\documentclass[twocolumn,showpacs,preprintnumbers,amsmath,amssymb]{revtex4}

\usepackage{graphicx}
\usepackage{dcolumn}
\usepackage{bm}

\newcommand{\be}{\begin{equation}}
\newcommand{\ee}{\end{equation}}
\newcommand{\bea}{\begin{eqnarray}}
\newcommand{\eea}{\end{eqnarray}}

\newcommand{\integer}{\relax{\rm I\kern-.18em N}}

\begin{document}

\title{$D$ mesons and charmonium states in hot pion matter}
\author{C. Fuchs$^1$}
\author{B.V. Martemyanov$^{1,2}$}
\author{Amand Faessler$^1$}
\author{M.I. Krivoruchenko$^{1,2}$}
 \affiliation{$^1$ Institut f\"{u}r Theoretische Physik$\mathrm{,}$
Universit\"{a}t
T\"{u}bingen$\mathrm{,}$,
Auf der Morgenstelle 14$\mathrm{,}$D-72076 T\"{u}bingen, Germany\\
$^2$ Institute for Theoretical and Experimental Physics,
B. Cheremushkinskaya 25, 117259 Moscow, Russia}
\date{}
\begin{abstract}
We calculate the in-medium $D$ meson self-energies in a hot pion
gas induced by resonance interactions with pions. The appropriate
resonances in the {\it s},~ {\it p} and {\it d} waves of the
$D$ meson-pion  pair are represented by low lying scalar, vector
and tensor $D^*$ mesons. At temperatures around 200 MeV
the D-meson mass drops by $30~ \rm {MeV}$ and the scattering width
grows up to $60~ \rm {MeV}$. Similar medium effects are
found for the $D^*$ vector mesons. This opens and/or enhances the
decay and/or dissociation channels of the charmonium states $\Psi^\prime$,
$\chi_c$ and   $J/\Psi$ to $D \bar D,~D^* \bar D,~D \bar D^*
,~D^* \bar D^*$ pairs in pion matter.

\end{abstract}
\maketitle
Since the conjecture of the dissociation of charmonium states
$J/\Psi,~\Psi^\prime ,\dots$ in a quark-gluon plasma (QGP)
due to color screening \cite{matsui86} $J/\Psi$ suppression
was considered as one of the  key signals for the creation of a 
QGP phase in heavy ion reactions. An anomalous $J/\Psi$ suppression 
has possibly been seen in central $Pb+Pb$ reactions at CERN Super Proton Synchrotron (SPS) 
\cite{na50}. However, the interpretation of data is strongly influenced by the
final state interactions of charmonium states in a hadron gas.
The dissociation of  $J/\Psi$'s due to pion scattering
$J/\Psi + \pi \rightarrow D{\bar D},D^*{\bar D},D{\bar D}^*,
D^*{\bar D}^*$ has extensively been discussed
within the comover scenario \cite{comover}, 
for a review see \cite{vogt99}. The $J/\Psi$ yield at
Relativistic Heavy Ion Collider (RHIC) 
will be measured by the PHENIX
Collaboration \cite{phenix04}. On the other hand, recent lattice QCD calculations
indicate that the $J/\Psi$ might exists as a bound state even
at temperatures above the critical temperature for a  QGP phase
transition \cite{lattice}.

Direct in-medium modifications of charmonium states, i.e.
mass shifts and elastic collisional widths are expected to be
small compared to those of the $D$ mesons \cite{arata}.
Indirect modifications of charmonium state can, however, appear due to medium
modifications of the $D$-meson states. With the lowering of $D$ masses  the
inelastic collisional width of charmonium states (due to
dissociation to open charm) or, in some cases, even the ordinary width
(due to on-shell decays to open charm) is increased
or is appeared as a result of the lowering of corresponding
thresholds \cite{friman}. These facts are important for both,
the production and the survival probability of $J/\Psi$  mesons.
About  $20\%$, $14\%$ and $8\%$ of $J/\Psi$'s are produced
in $pN$ interaction at $300~ \rm {GeV/c}$ via the decays of the
excited charmonium states $\chi_1$, $\chi_2$ and $\Psi^\prime$,
respectively \cite{Anton}. A dissociation of these
states in the medium
can approximately reduce the $J/\Psi$-production rate by a factor of two.
Moreover, the shift of the thresholds for the $D$-meson
channels in $J/\Psi$-$\pi$ collisions increases the $J/\Psi$-dissociation 
cross section and reduces the $J/\Psi$-survival probability.

$D$ mesons have an obvious analogy to $K$ mesons replacing
strange by  charm quarks. Medium modifications of
$D$ mesons and charmonium states have been considered 
by several authors. However, similar as for the kaons the investigation of 
D-meson mass shifts was mainly restricted to the case of nuclear
matter \cite{weise99,digal,sibirtsev02,brat2}.
In ultra-relativistic heavy-ion collisions
at CERN SPS or RHIC energies the hadronic
environment is, however, baryon dilute and meson rich \cite{bravina02}.
Thus, in the present work we investigate
in-medium properties of $D$ and $D^*$ mesons and the resulting 
consequences for charmonia in hot pion matter. Previous investigations
at finite temperature concentrated to large extent on the determination 
of charmonium dissociation cross sections
\cite{comover,vogt99,hashimoto86,wong00,ko,burau,ivanov04} which are, 
unfortunately, burdened by large model uncertainties.  

The strangeness sector has been investigated theoretically in  \cite{MFFK}
(see also references therein) and it has been found that kaons acquire a negative mass shift
and a substantial broadening which could  explain the enhanced 
$\phi \rightarrow \mu^{+}\mu^{-} $ decay \cite{NA50} relative to the 
$\phi \rightarrow K\bar{K}$ decay \cite{NA49} possibly seen in heavy ion 
collisions at CERN SPS (see although Refs.\cite{Jacak,Filip} on possible
explanation of the effect) 
and the  $\phi \rightarrow e^{+}e^{-} $ to
 $\phi \rightarrow K\bar{K}$ enhancement possibly seen in heavy ion collisions at
RHIC \cite{phenix}.    

The self-energy $\Sigma$ of a $D$ meson in a pion gas is 
to leading order in pion density determined by the 
invariant $D\pi$-forward scattering amplitude $A$ 
\be
\Sigma = -\int (A^{+}dn_{s\pi ^{+}}+A^{0}dn_{s\pi ^{0}}+A^{-}dn_{s\pi ^{-}})~.
\label{sigma}
\ee
Scalar $n_{s\pi}$ and vector pion densities  $n_{v\pi}$ 
($dn_{s\pi }=dn_{v\pi }/(2E_{\pi }))$ are given by 
Bose-Einstein distributions at fixed temperature $T$
\be
dn_{v\pi }=\frac{d^{3}p_{\pi }}{(2\pi )^{3}}\left( {\rm exp}(\frac{E_{\pi
}-\mu _{\pi }}{T})-1\right) ^{-1}~.
\ee
Here $\mu _{\pi ^{+}}=-\mu _{\pi ^{-}}$ is the $\pi ^{+}$ chemical
potential, $\mu _{\pi ^{0}}=0$. In isotopically symmetric pion 
matter which is expected to be formed in ultra-relativistic  
heavy-ion collisions at RHIC energies the chemical potentials 
vanish ($\mu _{\pi ^{\pm}}=0$). Hence all pions 
(negative, neutral and positive) are equally distributed.

In an isotopically symmetric pion gas $D^+$ and $D^0$ obtain  
identical self-energies due to isospin symmetry and these 
are further equal to the $D^-$ and $\bar D^0$ self-energies due to
charge conjugation symmetry. Thus we can restrict ourselves in the 
following to the discussion of the $D^+$ self-energy. In  
this context it should  be noticed that $D$ mesons show a close 
analogy to kaons. In both cases a mass splitting between $K$ and ${\bar K}$, 
respectively   $D$ and ${\bar D}$ occurs at finite nuclear density 
\cite{sibirtsev02} whereas in isotopically symmetric pion matter kaons 
and antikaons obtain equal in-medium modifications \cite{MFFK}. 

 The interaction of $D$ mesons with pions or
 more generally, the interaction of charmed heavy-light pseudoscalar 
 and vector mesons with light pseudoscalar mesons has been already discussed
 in the literature. A local 4-particle interaction motivated by
 chiral symmetry and by the similarity of kaons and $D$ mesons was 
introduced in Ref. \cite{lutz}. This tree level interaction has 
further been iterated for $s$-wave scattering \cite{lutz} and the 
 poles of the amplitudes in specific channels were identified 
 with recently observed scalar and axial $D^*_s$ mesons 
 \cite{BABAR,CLEO}. The tree level  $D^+\pi$-scattering amplitude 
reads \cite{lutz}
 \bea
 A(D^+\pi^+ \rightarrow D^+\pi^+)& = &-\frac{1}{4 F^2} (s-u)\nonumber\\
 A(D^+\pi^0 \rightarrow D^+\pi^0)& = &0\\
 A(D^+\pi^- \rightarrow D^+\pi^-)& = &\frac{1}{4 F^2} (s-u)\nonumber~,  
\eea 
where $F\approx 93~ \rm {MeV}$ is the pion decay constant. This result can 
be presented by isospin 1/2 and isospin 3/2  amplitudes 
 \bea 
 A_{1/2}& = &\frac{1}{2 F^2} (s-u)\nonumber\\
A_{3/2}& = &-\frac{1}{4 F^2} (s-u)~.
\eea
The first amplitude $A_{1/2}$  is positive and attractive,
the second amplitude $A_{3/2}$  is negative and repulsive.
Iterating the $A_{1/2}$ amplitude to all orders  in ladder
 approximation leads to a large value for the $T$ matrix (at least in $s$ waves) 
and a pole corresponding to a $D\pi$ resonance.  The  $A_{3/2}$
 amplitude after iteration  decreases \cite{lutz}.
 The tree level amplitudes contain only $s$  and $p$ waves.
Not discussing the reliability of such an approach
in what follows we will not rely on these tree level amplitudes but
instead will take only the observed $s$- , $p$-  and $d$- wave resonances in the 
$D\pi$ system into account. 
For orbitally excited mesons we take the experimental data
from the BELLE Collaboration \cite{BELLE} (see Fig.1)
where for the first time
all four orbitally excited $D$-meson states have been observed
 simultaneously. The information on the very narrow vector mesons $D^*$
($p$-wave resonances close to threshold)
is taken from Particle Data Group\cite{PDG}.

\begin{figure}[!htb]
\begin{center}
\includegraphics[width=.48\textwidth]{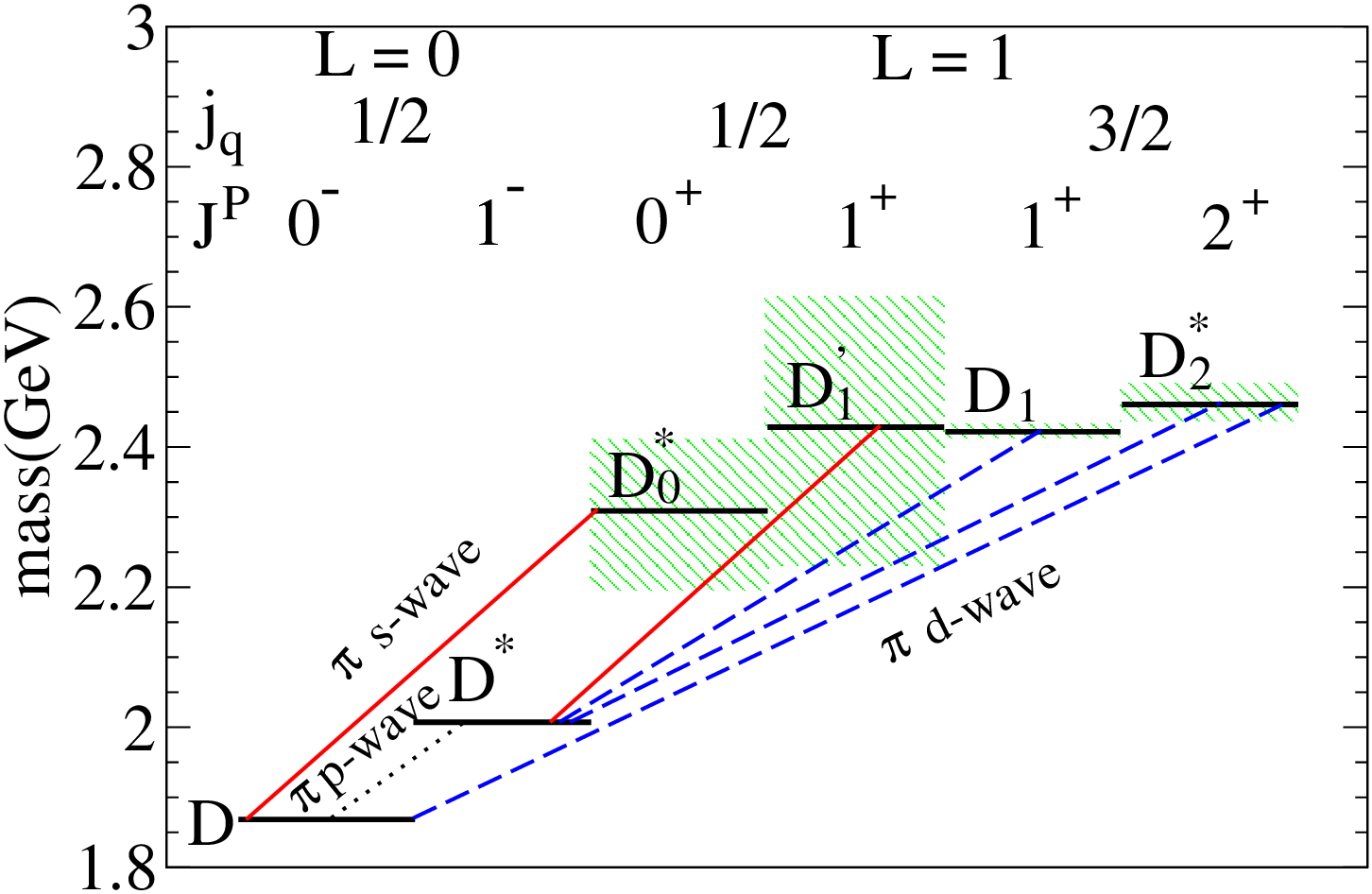}
\caption{(Color online) Levels of $D$-meson excitations from Ref.  \cite{BELLE}.}
\label{levels}
\end{center}
\end{figure}

\begin{table}[tbp]
\caption{Excited $D$-meson states which are taken into account as
resonances in $D\pi$ system. }
\label{labl2}
\begin{center}
\begin{tabular}{lcc}
\hline
Resonance & mass (MeV)  & width (MeV) \\ \hline
$D^*$ & 2008.5 & $\approx$ 0.1 \\
$D^*_0$ &$ 2308\pm 60$ & $276\pm 99$ \\
$D^\prime_1$ &$ 2427\pm 61$ &$ 384^{+201}_{-169}$\\
$D_1$ &$ 2421.4\pm 2.7$ &$ 23.7\pm 6.9$ \\
$D^*_2$ &$2461.6\pm 5.9$  &$ 45.6\pm 12.5$\\
\hline
\end{tabular}
\end{center}
\end{table}
For  resonance scattering we include only the
isospin 1/2 resonances which leads to
\bea
 A(D^+\pi^+ \rightarrow D^+\pi^+)& = &0\nonumber\\
 A(D^+\pi^0 \rightarrow D^+\pi^0)& = &\frac{1}{3} A_{1/2}\\
 A(D^+\pi^- \rightarrow D^+\pi^-)& = &\frac{2}{3} A_{1/2}\nonumber~
\eea
and the sum of these amplitudes that enters the thermal average is
equal to $A_{1/2}$.  For the forward resonance amplitude $A_{1/2}$ we use
the relativistic Breit-Wigner form
\be
A_{1/2} =  \sum_{j=0,1,2} \frac{8\pi\sqrt{s}}{k}
\frac{(2j+1)}{(2j_1+1)(2j_2+1)}
\frac{-\sqrt{s}\Gamma_j^{D\pi}}{s-M_j^2 +i\sqrt{s}\Gamma_j^{tot}}~,
\ee
where $j=0,1,2$ corresponds to the $s-,p-$ and $d-$ wave resonances
$D^*_0,~ D^*$ and $D^*_2$ in the $D\pi$ system with masses
$M_j$, partial and total widths $\Gamma_j^{D\pi}$ and $\Gamma_j^{tot}$
, respectively;
$j_1 = j_2 = 0$ are the spins of the $D$ and the pion, respectively,
and $k$ is the  c.m. momentum. 
The energy dependence of the widths is regulated by the specified 
partial wave
\be
\Gamma_j^{D\pi}=\left(\frac{k}{k_0}\right)^{2j+1}\frac{M_j^2}{s}\Gamma_{j~0}^{D\pi}~,
\ee where subscript '0' refers to on mass shell decay widths.
When the same amplitude is evaluated for
the $j_1 = 1$ $D^*$ meson, the sum runs over the  resonances
$D_{1}^\prime ,~ D_1,~D^*_2$ and the ground state $D$ meson.
 For the branching of the tensor resonance
we take the world average
$\Gamma(D^*_2\rightarrow D\pi):\Gamma(D^*_2 \rightarrow D^*\pi)\approx 2:1$.
\begin{figure}[!htb]
\begin{center}
 \includegraphics[width=.48\textwidth]{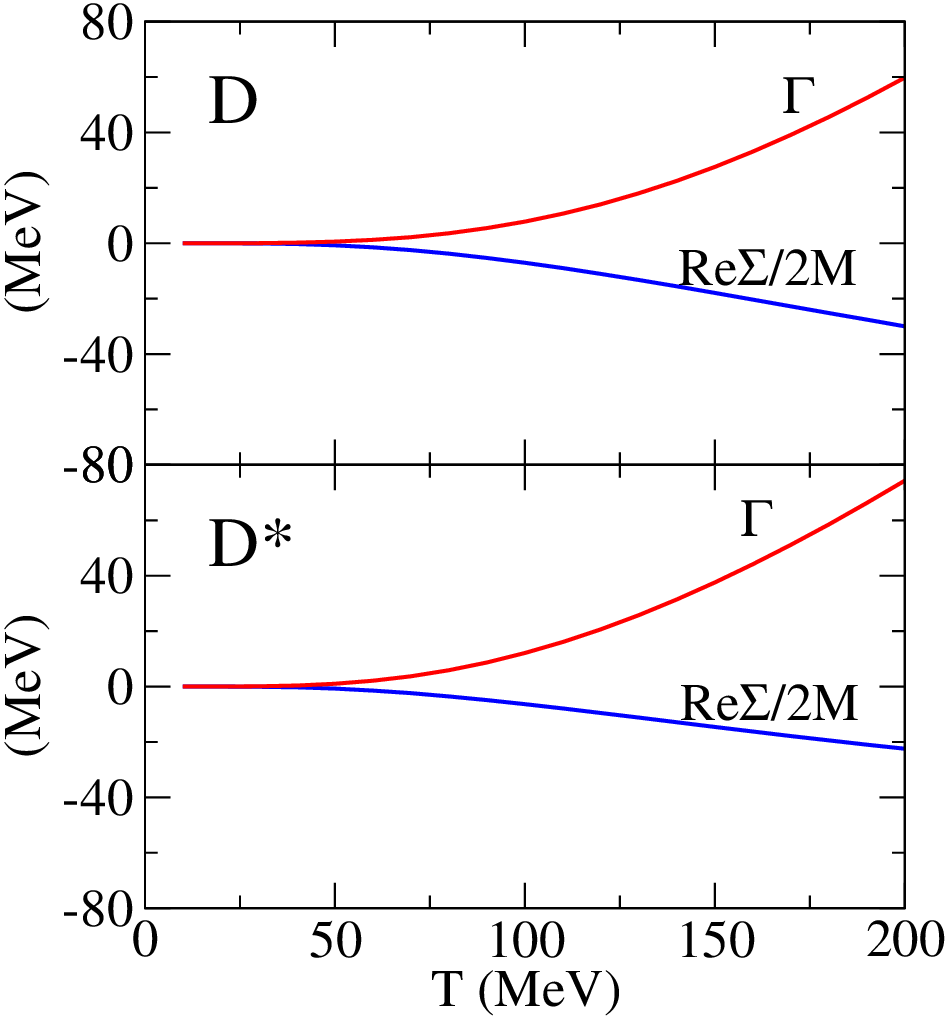}
\caption{(Color online) The scattering width $\Gamma=-Im\Sigma/M$ and real part of 
self-energy $Re\Sigma/2M$
 as a function of temperature $T$ are shown for
for $D$ mesons (top) and  $D^*$ mesons (bottom) at rest in pion matter .}
\label{dmesmas}
\end{center}
\end{figure}
Now the $D$, respectively the  $D^*$  self-energy in a pion gas at rest
 (\ref{sigma}) is obtained by integration over the pion distribution
which yields the corresponding modifications of the meson properties:  
the scattering width $\Gamma$ and mass shift $Re\Sigma/2M$.
The contribution of the narrow $D^*$ resonance with a width of about
$100~ \rm {KeV}$ to the $D$ meson medium modifications are of the order
of the $D^*$ width and can be neglected (if one takes the in-medium 
modification of the $D^*$ into account, this is no more the case as will 
be seen later on). The contributions of the 
scalar and tensor $D^*$ mesons are, however, sizable and 
shown in Fig. 2 as a function of the temperature.  
Medium modifications of the vector $D^*$ mesons which are of 
similar magnitude arise from the coupling to axial and tensor
mesons, see also Fig.2.

For further applications to charmonium states the off-shell properties
of $D$ and  $D^*$ mesons, namely the spectral functions will be important.
Spectral function is defined by the off-shell self-energy $\Sigma(m^2,|{\bf p}|)$
as follows
\be
\rho(m^2,|{\bf p}|)=\frac{1}{\pi}\frac{-Im\Sigma}
{(m^2-M^2-Re\Sigma)^2+(Im\Sigma)^2}~.
\ee
In medium spectral function depend not only on the invariant mass
squared of the particle but also on its momentum. For $D$ and  $D^*$ mesons
at rest in the hot pion medium with temperature $T = 200~\rm {MeV}$ 
the spectral functions are shown on Fig.3 as the functions of the invariant mass.
At other momenta typical for the decays of the charmonium states (see below)
they don't change significantly because the velocities of $D$ and  $D^*$ mesons
are small compared to the thermal velocities of pions.

\vspace{1.2 cm} 
\begin{figure}[!htb]
\begin{center}
 \includegraphics[width=.48\textwidth,height=.6\textwidth]{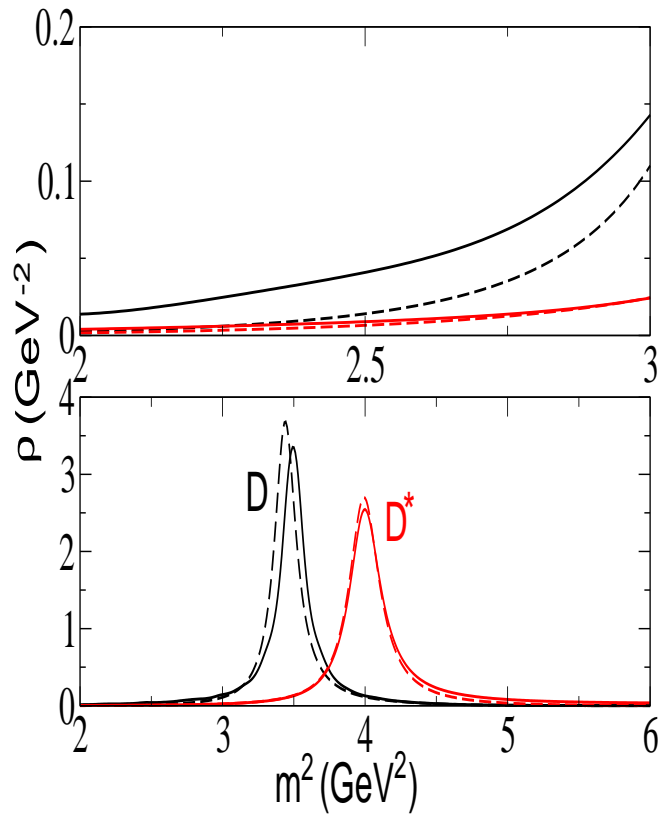}
\caption{(Color online) The in-medium spectral functions of $D$ and $D^*$ mesons
at rest in the hot pion gas with temperature $T = 200~\rm {MeV}$.
The dashed and solid lines for the in-medium $D,D^*$-meson spectral functions
 correspond to the lowest order calculation 
and to the first order self-consistent iteration of  
the in-medium spectral functions.} 
\label{dmesspecfun}
\end{center}
\end{figure}

For $D,D^*$ mesons we calculated the spectral functions both, 
with vacuum (dashed lines)
and in-medium (solid lines) spectral functions of $D^*,D$ mesons. 
In the latter case which is the first order self-consistent 
iteration, the vacuum propagators of $D$ and $D^*$ in (6) are replaced by the 
corresponding in-medium propagators in order to obtain the iterated 
in-medium results
\be
\frac{1}{s-M_{D(D^*)}^2}
\rightarrow
\frac{1}{s-M_{D(D^*)}^2 -\Sigma_{D(D^*)}}~.
\ee
 The medium modification of the $D$ meson has 
thereby a feedback on the $D^*$ since it modifies the corresponding 
decay channel. The difference between the lowest order and the first 
iteration is significant at small invariant masses of the $D,D^*$ mesons
which is important for the charmonium decays.

\vspace{1.2 cm}
\begin{figure}[!htb]
\begin{center}
 \includegraphics[width=.48\textwidth,height=.48\textwidth]{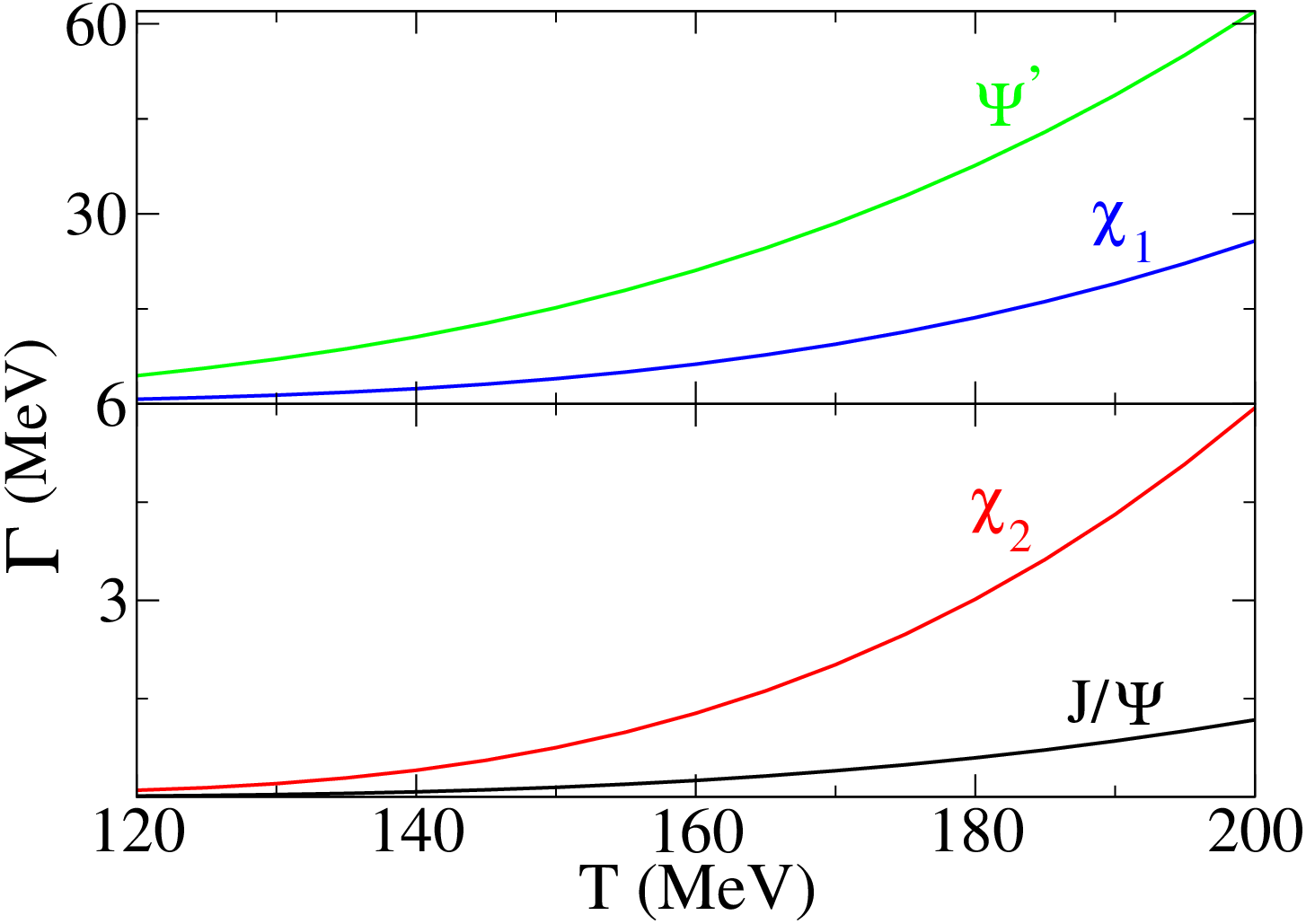}
\caption{(Color online) In-medium width of $J/\Psi,~ \chi_1,~\chi_2$ and $\Psi^\prime$
mesons.} 
\label{jwidth}
\end{center}
\end{figure}

With this information, 
one is able to evaluate the  dissociation widths of the charmonium states 
$J/\Psi ,~ \chi_1,~\chi_2,~\Psi^\prime $ in the pion medium.
 Let us take into account the fact 
that in the medium $D,~D^*$ mesons are not particles but 
resonances with  finite collisional widths. 
Then the $J/\Psi \rightarrow D\bar D,~ D\bar D^*
,~ D^*\bar D,~ D^*\bar D^*$ decay can occur subthreshold 
and thus account for  the reactions 
$J/\Psi \pi\rightarrow D\bar D,~ D\bar D^*,~ D^*\bar D,~ D^*\bar D^*$ 
which are open in pion matter. 
The  width
$\Gamma(J/\Psi \rightarrow D\bar D)$ can be easily found
\begin{eqnarray}
&&\Gamma(J/\Psi \rightarrow D\bar D) =\int
\frac{g^2_{J/\Psi D\bar D}}{3\pi M^2_{J/\Psi}}p^{3}\nonumber \\
&&\times \rho_D(m_1^2,p)\rho_{\bar D}(m_2^2,p)dm_1^2 dm_2^2 ~,\nonumber
\label{jwidtheq}
\end{eqnarray}
where $p$ is c.m. momentum for the decay of 
${J/\Psi}$ meson to $D$ and  $D^*$ mesons  with masses $m_1$ and 
$m_2$, respectively and $g_{J/\Psi D\bar D}$ is
the $J/\Psi D\bar D$ coupling constant. Taking 
the value of $g_{J/\Psi D\bar D} = 7.8$ from ref. \cite{friman} 
one obtains a dissociation width of 
$\Gamma(J/\Psi \rightarrow D\bar D) =
 0.54~ \rm {MeV}$ at a temperature of $T = 200~ \rm {MeV}$.
 The other decay channels ($D\bar D^*~ D^*\bar D,~ D^*\bar D^*$) can be
 treated analogously. To do so, we use the coupling constants
 $$ g_{J/\Psi D^*\bar D^*}=g_{J/\Psi D\bar D},~~~~~~ g_{J/\Psi D\bar D^*}
 =\frac{g_{J/\Psi D\bar D}}{M_D}~$$
 in combination with appropriate phenomenological vertices \cite{ko}.
 In total one obtains thus at a temperature 
 of $T = 200 ~\rm {MeV}$ a  $J/\Psi$ collisional width of 
 $1.15~\rm {MeV}$ (see Fig.4).  

These values can now be compared
 to the collisional width $\Gamma(J/\Psi \pi \rightarrow D\bar D 
 ,~ D\bar D^*,~ D^*\bar D,~ D^*\bar D^*) 
 = 5\div 14~ \rm {MeV}$. The later one was obtained by multiplying the 
 average cross section for the  $J/\Psi$ dissociation through 
pions evaluated at the same temperature and equal to $<\sigma^{\pi J/\Psi}
 v> \approx 0.75\div 2~ \rm {mb} $ \cite{ko} 
 ($v$ being the relative velocity)
 with the corresponding pion density.  The order of magnitude 
agreement between  the  $J/\Psi$ subthreshold decay width to $D$ mesons 
and the  
$J/\Psi \pi$ dissociation width is not surprising because
 both effects are of common
nature. In fact, the $J/\Psi$ dissociation through pion capture by one of the
$D$ mesons is connected to the $D$ mesons collision widths. In the 
calculation of the $J/\Psi$ decay rate the $D$ mesons collision width
appears in the denominator of the Breit-Wigner amplitude which, 
due to the subthreshold character of the $J/\Psi$ decay, can be 
expanded in powers of the pion density entering the $D$-meson collision
 width. To lowest order 
the result coincides with the $J/\Psi$ dissociation rate through 
one-pion capture by one of the $D$ mesons. However,  
the inclusion of  realistic spectral functions turns out to 
reduce the  $J/\Psi$ dissociation. 
So,  the results of previous calculations\cite{ko} 
differ from the present ones because some contribution
(contact 4-particle interaction) is not considered in our approach
(see diagram 1c in Fig. 1 in \cite{ko} ) and because our amplitude is
closer to the unitarity constraint.
In this context it should 
be mentioned that there appear additional 
dissociation mechanisms at the quark level \cite{burau,ivanov04,Marina}, e.g. 
$s$-channel box diagrams, which cannot be
expressed in terms of effective meson degrees of freedom and may 
lead to an additional increase of the width.

The in-medium width of the excited $\Psi^\prime$ is estimated analogously 
assuming the same $g_{\Psi^\prime D\bar D}$ coupling constant as for the 
$J/\Psi$. Since the  $\Psi^\prime$ lies only $52~\rm {MeV}$ below 
the $ D\bar D$ threshold in free space its in-medium width  
is about $50$ times larger than that of the $J/\Psi $. 
Finally, the in-medium widths of the $\chi_1,~\chi_2$ states were 
estimated using the 
phenomenological vertices $g M_D \chi_{\mu} (D_{\mu} \bar D +\bar 
D_{\mu} D)$ and $g M_D \chi_{\mu\nu}  D_{\mu} \bar D_{\nu}$,~
$g/ M_D \chi_{\mu\nu} \partial_{\mu} D \partial_{\nu}\bar D$
with $g=g_{J/\Psi D\bar D}$, respectively. As seen from Fig.4 
 $\chi_1,~\chi_2$ mesons also receive a substantial width in pion matter.

In conclusion, the $D$ meson self-energies in pion 
matter have been determined to leading order in density taking thereby 
resonances in the $D\pi$ amplitude into account. The 
resonances in $s$, $p$ and $d$ waves of~$D$ meson-pion system were
represented by low lying scalar, vector and tensor $D^*$ mesons which 
have been observed experimentally. This allows to determine the 
$D$ self-energy in a model independent way. 
At a temperature around $200~ \rm {MeV}$ 
the $D$-meson mass is reduced by about $30~ \rm {MeV}$ and the 
scattering width is  about $60~ \rm {MeV}$. 
Similar medium modifications were found for $D^*$ vector mesons.
Consequently, the widths
of the decay and dissociation channels of the charmonium states $\Psi^\prime$, 
$\chi_c$ and   $J/\Psi$ to $D \bar D,~D^* \bar D,~D \bar D^*
,~D^* \bar D^*$ pairs are  
enhanced (from $\Gamma_{J/\Psi} \simeq 1.15$ MeV to 
$\Gamma_{\Psi^\prime} \simeq 62$ MeV at $T\simeq 200$ MeV). 
As a consequence, feeding of  $J/\Psi $ states from excited charmonium 
states ceases  in a hot pion gas which characterizes the hadronic 
final state in high energetic heavy ion reactions in good approximation.

 Hence, $D$ mesons modifications in hot pion medium are 
important  for the production of $J/\Psi$ 
 during the fireball expansion in heavy ion 
reactions. The back reactions of $J/\Psi$ 
formation in $D$ mesons collisions from charmed meson rich medium  
become important  for corresponding transport simulations.

This work is supported by RFBR grant No.
06-02-04004 and DFG grant No. 436 RUS 113/721/0-2. 
M.I.K. and B.V.M.  acknowledge the kind hospitality at the University
 of T\"{u}bingen.


\end{document}